\documentclass[showpacs,twocolumn,prl]{revtex4}
\usepackage{amsmath}
\usepackage{amsfonts}
\usepackage{graphicx}
\usepackage{bm}
\begin{document}

\title{Spin relaxation and decoherence of two-level
systems}

\author{X. R. Wang}
\affiliation{Physics Department, The Hong Kong University of
Science and Technology, Clear Water Bay, Hong Kong, China}
\author{Y. S. Zheng}
\affiliation{Physics Department, The Hong Kong University of
Science and Technology, Hong Kong, China}
\affiliation{Physics Department, Jinlin University, Changchun, China}
\author{Sun Yin}
\affiliation{Physics Department, The Hong Kong University of
Science and Technology, Hong Kong, China}

\date{\today}
\begin{abstract}
We revisit the concepts of spin relaxation and spin decoherence of
two level (spin-1/2) systems. From two toy-models, we clarify two
issues related to the spin relaxation and decoherence: 1) For an
ensemble of two-level particles each subjected to a different
environmental field, there exists an ensemble relaxation time
$T_1^*$ which is fundamentally different from $T_1$. When the
off-diagonal coupling of each particle is in a single mode with the
same frequency but a random coupling strength, we show that $T_1^*$
is finite while the spin relaxation time of a single spin $T_1$ and
the usual ensemble decoherence time $T_2^*$ are infinite. 2) For a
two-level particle under only a random diagonal coupling, its
relaxation time $T_1$ shall be infinite but its decoherence time
$T_2$ is finite.
\end{abstract}
\maketitle
\begin{figure}[htbp]
 \begin{center}
\includegraphics[width=7.cm, height=5.5cm]{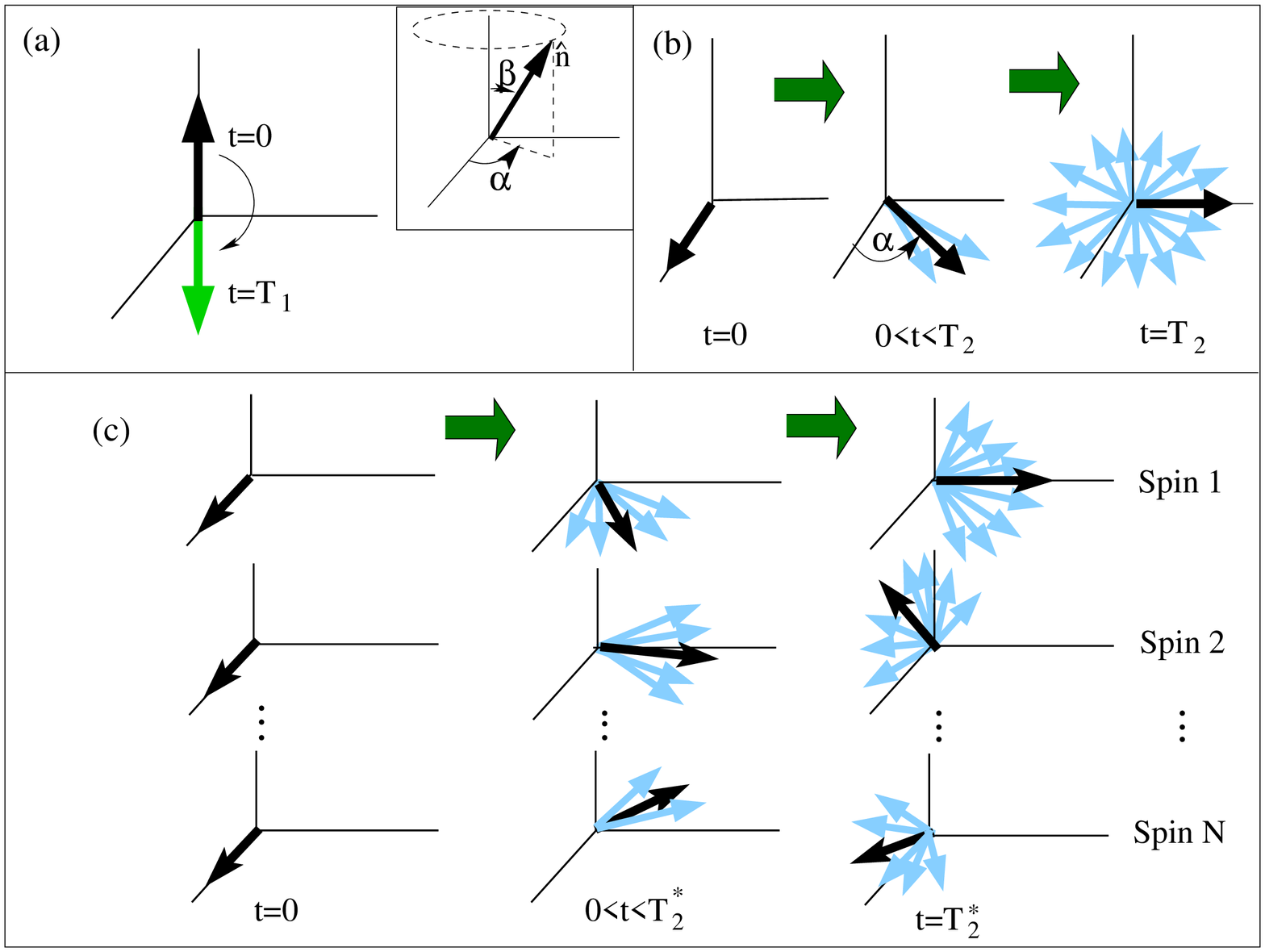}
 \end{center}
\caption{(Color) An illustration of $T_1$, $T_2$ and $T_2^*$. (a)
$T_1$ is the typical time for a spin jumps from the upper level (up
spin) to the lower level (down spin). Inset: The polar angle $\beta$
and the azimuthal angle $\alpha$ of a spin along unit direction
$\hat{n}$. (b) The precession of spin in xy-plane. At $t=0$, the
spin points to x-direction. Due to its irregular precession, the
spin direction in a later time may be in a range of directions
denoted by grey (blue) arrows. The spin loses its direction
completely at $T_2$, called the decoherence time. (c) For an
ensemble of spins, all spins point to the x-direction at the
beginning. As time goes on, different spins will precess to
different directions due to the fluctuating field is different at
different places. After $T_2^*$, spins point to all directions all
the time. \label{fig1}}
\end{figure}

Spin relaxation and decoherence is of central importance in quantum
computation and spintronics\cite{loss1,kane,dassarma}. Two
phenomenological quantities, $T_1$ and $T_2$ known as the spin
relaxation time and spin decoherence time for a two-level (spin-1/2)
system, were introduced to describe the spin dynamics under the
influence of an environment or external field in early
theory\cite{bloch,torrey}. As illustrated in Fig. \ref{fig1}(a), a
spin in its upper level, denoted by $|+\rangle$, may jump to its
lower energy level, denoted by $|-\rangle$, due to spin-environment
coupling. $T_1$ is defined as the average transition time. A quantum
spin can also be in a superposition state, say
$(|+\rangle+|-\rangle)/\sqrt{2}$ initially. Without the
environmental interaction, the system undergoes a constant
precession with a frequency $\omega=E/\hbar$, where $E$ is the
energy difference between $|+\rangle$ and $|-\rangle$. Under the
environmental influence, the phases of the expansion coefficients in
the superposition state will become random, and their relative
phases, which describe the precession angle $\alpha$ in Fig.
\ref{fig1}(b), will be completely undetermined (decoherence). The
time for the superposition state to lose its coherence is called
$T_2$. $T_1=T_2$ is believed to be true when the spin is isotropic
while $T_2\leq 2T_1$ holds in general\cite{weiss,yafet}.
\par
In recent years, it has been realized that another quantity,
$T_2^*$, is important for a large number of experiments that measure
the decoherence of an ensemble of spins rather than a single
spin\cite{glazman,hu}. Illustrated in Fig. \ref{fig1}(c), in
addition to the uncertainty in the precession angle of a single spin
at a given time, there will be an uncertainty in the relative
precession angles among different spins due to the spatial
inhomogeneity. This additional decoherence source leads normally to
$T_2^* \le T_2$, within which $T_2^*$ may deviate notably from
$T_2$. For example, $T_2^*$ of electron spins in quantum dots was
shown to be several orders smaller than $T_2$\cite{glazman}.
\par

Our current understanding\cite{dassarma} of $T_1$, $T_2$ and $T_2^*$
is more accurate and much deeper than that of the early years\cite
{legget}. Its applications to various systems, especially those of
nanostructures are of current focus. However, there are still many
vague conceptions or even misconceptions about the description,
origins and mechanisms of these quantities. In this letter, we
revisit the concepts of the spin relaxation and decoherence of
two-level systems. We shall argue that, besides an ensemble
decoherence time $T_2^*$, an ensemble relaxation time $T_1^*$ is
also necessary to interpret experimental results. We show, through a
toy-model calculation, that $T_1^*$ distinguishes itself from $T_1$.
In the case of our study, the total polarization of the system
decreases with time while each spin undergoes a well-defined
periodic oscillation. Thus, there is a finite pseudorelaxation time
$T_1^*$ for the whole system, but an infinite $T_1$. We investigate
also the effects of a random diagonal coupling on spin relaxation
and decoherence. We show that both slow and fast fluctuation are
harmful to $T_2$ while its effect on $T_1$ must go through the
off-diagonal coupling.

\par
{\bf {\it Formulation of the dynamics of a two-level system.}} To
mimic the motion of a two-level (spin-1/2) system coupled to an
environment, we assume that the system may jump from one state into
the other through the coupling, or the system in a superposition
state may obtain random phases to its expansion coefficients. The
Hamiltonian of the system can be properly described by a $2\times 2$
matrix,
\begin{equation}
H=\frac{E}{2}\sigma_z+\bm{K}(t)\cdot\bm{\sigma}=\left[
\begin{array}{cc}\frac{E}{2}+K_z&K_x-iK_y\\K_x+iK_y&
-\frac{E}{2}-K_z\end{array}\right]. \label{hamiltonian}
\end{equation}
$\bm{\sigma}=\sigma_x\hat{\bm{x}}+\sigma_y\hat{\bm{y}}+
\sigma_z\hat{\bm{z}}$ is the Pauli operator.
The environment-system interaction is approximated by a
randomly fluctuating field $\bm{K}(t)=K_x\hat{\bm{x}}+K_y\hat
{\bm{y}}+K_z\hat{\bm{z}}$, which could originate from the
interactions of the spin with phonons, photons or with other
surrounding spins. In general, the field is correlated both
in time and in space. Its covariance measures the strength
of the coupling which should also depend on the temperature
besides of other factors. How to obtain the field from a
microscopic Hamiltonian of both system and the environment is
an interesting problem in both physics and chemistry\cite{legget}.
For example, it has been shown\cite{nazarov2} that electron-phonon,
spin-orbit, and/or hyperfine interactions for electron spins in a
quantum dot are equivalent to off-diagonal couplings. We will see
that the diagonal coupling $K_z(t)$ and the off-diagonal coupling
$K_{x(y)}(t)$ play different roles in determining $T_1$ and $T_2$
respectively. Without losing the generality, the energies of the
upper and the lower levels ($|\pm\rangle$) are assumed to be
$\pm E/2$, respectively, in the absence of the fluctuating field.
The energy may be due to the Zeeman interaction for a spin in a
constant magnetic field.
\par
Express the general wave function of our two-level system as
$\psi(t)=a(t)|+\rangle+b(t)|-\rangle$, the equations for the
expansion coefficients $a(t)$ and $b(t)$, derived from the
time-dependent Schr\"{o}dinger equation, are (set $\hbar=1$)
\begin{eqnarray}
&&i{da\over dt}=[E/2+K_z]a+[K_x-iK_y]b,\nonumber\\
&&i{db\over dt}=-[E/2+K_z]b+[K_x+iK_y]a. \label{schrodinger}
\end{eqnarray}
The issues are to find out the spin relaxation and decoherence time
for a given fluctuating field. Spin-relaxation time $T_1$ can be
obtained by examining how $\langle|a(t)|^2\rangle$ changes with time
under an initial condition, where $\langle \ldots \rangle$ denotes
the average over the system state. $T_1$ is the characteristic decay
time extracted from $\langle|a(t)|^2\rangle$.
\par
To evaluate $T_2$ or $T_2^*$, it is useful to notice that a spin
state along a particular direction $\hat{n}$, specified by the polar
and azimuthal angles $\beta$ and $\alpha$ [inset of Fig. 1(a)], can
also be written as $\psi(t)=\exp(i\phi)[\cos(\beta/2)\exp
(i\alpha/2)|+\rangle+\sin(\beta/2)\exp(-i\alpha/2)|-\rangle$]. If
one can find $\alpha(t)$ by solving Eq.({\ref{schrodinger}), then
$T_2$ for a single spin or $T_2^*$ for an ensemble of spins
corresponds to the time at which the deviation of $\alpha$ equals
$\pi$, i.e., $\langle \Delta\alpha^2(T_2) \rangle=\langle\alpha
(T_2)^2\rangle-\langle \alpha(T_2)\rangle ^2=\pi^2 $.
\par

{\bf {\it Ensemble spin relaxation time $T_1^*$.}} In order to
demonstrate the necessity of introducing an ensemble spin relaxation
time $T_1^*$, we consider an ensemble of noninteracting spin-1/2
particles each of which is described by a Hamiltonian similar to
(\ref{hamiltonian}) with $K_z=0$, and $K_x+iK_y=Ke^{i\omega t}$.Then
Eq.(\ref{schrodinger}) can be solved analytically with solution
\begin{eqnarray}
a_k(t)=&&e^{-i(\omega/2)t}[a_{k}(0)\cos(\Omega
t)\nonumber\\
&&+i\frac{\Delta\omega a_k(0)-2K_kb_k(0)}{2\Omega}\sin(\Omega t)],
\nonumber\\
b_k(t)=&&e^{i(\omega/2)t}[b_{k}(0)\cos(\Omega
t)\nonumber\\
&&-i\frac{\Delta\omega b_k(0)+2K_ka_k(0)}{2\Omega}\sin(\Omega t)],
\label{solution1}
\end{eqnarray}
where$\Delta\omega\equiv \omega-E$ and
$4\Omega^2=4K_k^2+\delta\omega^2$.
\par
 To make following analysis simple, we shall first assume $\omega=E$
 for all spins, and $K=K_k>0$ a random real constant for $k$th spin.
 Thus, each spin is nothing but the famous Rabi problem at resonance.
For the kth spin, Eq. (\ref{solution1}) gives
$a_k(t)=e^{-i(E/2)t}[a_k(0)\cos(K_kt)-ib_k(0)\sin(K_kt)]$ and
$b_k(t)=e^{i(E/2)t}[b_k(0)\cos(K_kt)-ia_k(0)\sin(K_kt)]$. If one
measures the expectation value of $S_z$ of the $k$th spin defined as
$\langle S_{k,z}\rangle\equiv
|a_k(t)|^2\frac{1}{2}-|b_k(t)|^2\frac{1}{2}$ ($\hbar$is set to 1),
one finds $\langle
S_{k,z}\rangle=[|a_k(0)|^2-|b_k(0)|^2]/2\cos(2K_kt)+\Im[a_k(0)b_k^*(0)]\sin(2K_kt)$,
which oscillates periodically in time instead of decay. One then
should conclude $T_1=\infty$ as it should be. However, if we measure
$S_z$ of the whole system, we need to include contributions from all
spins. Since different spins have different K values, they will
oscillate with different periods, and the spins initially in phase
will be out of phase during the evolution, as is illustrated in Fig.
2. Thus, even all spins are fully polarized along z direction
initially, $S_z$ will decrease with time. If $K$ has a distribution
function of $P(K)=(2K/\sigma)e^{-K^2/\sigma}$, where $\sigma$
measures the width of the distribution, by using Eq.
(\ref{solution1}), we have $[\langle
S_k\rangle]\equiv(1/N)\sum_k\langle
S_{k,z}\rangle=1-\sqrt{\pi\sigma}te^{-\sigma
t^2}erfi(\sqrt{\sigma}t)$, where erfi(x) is the imaginary error
function, and $[\cdots]$ is an average over all spins. This time
dependence is plotted in the inset of Fig. 2. Thus, one finds the
polarization of the system relaxes after a characteristic time
$T_1^*=1/\sqrt{\sigma}$, although there is no relaxation for
individual spin. It is also interesting to evaluate $T_2$ and
$T_2^*$ of this toy model. Set $a_k(0)=b_k(0)=1/\sqrt{2}$ for all
$k$ and compare $a_k(t)$ and $b_k(t)$ with their expressions in
terms of the polar and azimuthal angles mentioned in the early
paragraphs, we find $\alpha_k(t)=Et$. All spins precess at the same
rate and in phase. Thus, both $T_2$ and $T_2^*$ are infinite. This
is a peculiar case in which each spin follows a deterministic
motion, and there is no relaxation and no decoherence. But if one
measures the polarization of the whole system, the system appear to
have a relaxation time $T_1^*$ that is fundamentally different from
$T_1$.

\par
In fact, one may also assume that $k$th spin has $\omega_k$ and
different level spacing $E_k$. Since $E_k$, $\omega_k$, and $K_k$
are constants for a given spin, $a_k(t)$ and $b_k(t)$ are
well-defined periodic functions according to Eq. (\ref{solution1}).
Thus its precession angle and expectation value of $S_{k,z}$, which
can be obtained after some algebras
\begin{align}
\langle
S_{k,z}\rangle=\frac{|a_k(0)|^2-|b_k(0)|^2}{2\Omega^2}[\Delta
\omega_k^2/4+K_k^2\cos(2\Omega t)],\nonumber
\end{align}
which oscillates periodically in time. Therefore, both $T_1$ and
$T_2$ are infinite. However, if one measures $S_z$ or $S_x-iS_y$
(which is a measure of precession) of the whole system, we need to
sum over all spins. This corresponds to a sum of many periodic
functions with different periods. Similar to the case of random
$K_k$ discussed early, it leads to a decay function, and $T_1^*$ and
$T_2^*$are finite because different spins precess with different
speeds in this case.

\begin{figure}[htbp]
 \begin{center}
\includegraphics[width=7.cm, height=5.5cm]{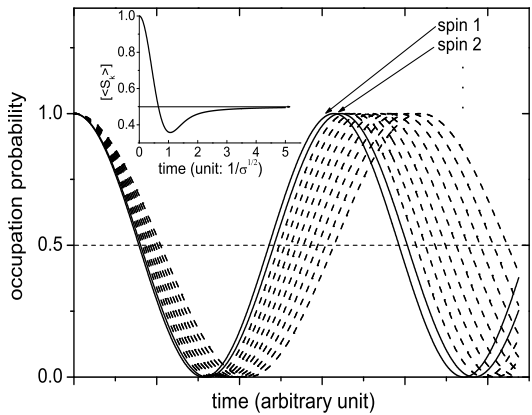}
 \end{center}
\caption{The time dependence of occupation probability of
different spin in its upper level. All spins are initially in
their upper levels. Due to the random coupling strength, their
Rabi frequencies are also random. Inset: The time dependence
of $[\langle S_{k} \rangle ]$. \label{fig2}}
\end{figure}
\par
Away from the resonance $\omega\ne E$, a spin initially at
the upper level will not evacuate completely from the
level. The occupation probability for a spin in the upper
level is given by the well-known Rabi solution
\begin{equation}
|a(t)|^2=1-A(\omega) \sin^2(\Omega t), \label{rabi}
\end{equation}
where $4\Omega^2=4K^2+(\omega-E)^2$, and $A(\omega)=4K^2/[4K^2+
(\omega-E)^2]$. Amplitude $A$ at the resonance takes the maximal
value 1. $A$ decreases fast when $\omega$ deviates from $E$.
Thus, the mode of the fluctuating field near the
resonance is expect to relax the spin more effectively.
We shall see that the diagonal coupling behaves differently.

\par
{\bf {\it Effects of diagonal coupling on $T_1$ and $T_2$.}} We
would like to discuss how a diagonal coupling affects the spin
decoherence and the spin relaxation. Consider a case in which $K_z$
in Eq. (\ref{hamiltonian}) is a randomly fluctuating field and
$K_x=K_y=0$, this is an exactly solvable problem because there is no
coupling between different levels, and $K_z$ causes only the energy
levels to move up and down randomly. The consequence is the
accumulation of a dynamical random phase for each level. The
solutions for the expansion coefficients are
$a(t)=a(0)\exp[-i\int_0^t(E/2+K_z(t'))dt']$ and $b(t)=b(0)
\exp[i\int_0^t(E/2+K_z(t'))dt']$, where $a$ and $b$ are separated
because of the absence of off-diagonal matrix elements. Thus, a spin
in either upper or lower level will not relax, and $T_1$ is
infinite. However, for a spin in superposition state, each of the
two expansion coefficients will obtain a random phase due to random
field $K_z$, leading to a random precession angle
$\alpha(t)=Et+2\int_0^tK_z(t')dt' $. Thus, the expectation of the
deviation of the precession angle is
\begin{equation}
\langle \Delta\alpha^2(t) \rangle=4\int_0^tdt_1dt_2\langle
K_z(t_1)K_z(t_2)\rangle. \label{deviation}
\end{equation}

For a stochastic field with autocorrelation,
\begin{subequations}
\begin{equation}
\langle K_z(t_1)K_z(t_2)\rangle={\Delta\over 2\tau_c}
\exp(-{|t_1-t_2|\over\tau_c}), \label{correlation}
\end{equation}
where $\tau_c$ and $\Delta$ are the correlation time and strength,
we have
\begin{equation}
\langle \Delta\alpha^2(t) \rangle=4\Delta
[t+\tau_c(e^{-\frac{t}{\tau_c}}-1)].
\end{equation}
\end{subequations}
Thus, the spin-decoherence time $T_2$ is given by
\begin{equation}
T_2+\tau_c[\exp(-T_2/\tau_c)-1]=\pi^2 /(4\Delta).
\label{T2}
\end{equation}
In the limit of $\tau_c \rightarrow 0$, the fluctuating field is
a white noise $\langle K_z(t_1)K_z(t_2)\rangle=\Delta
\delta(t_1-t_2)$, and the decoherence time is simply
$T_2=\pi^2/(4\Delta)$, inversely proportional to the
correlation strengthen of fluctuating field and independent from
the level spacing. Therefore, if the amplitude of the fluctuating
field increases $n$ times, $T_2$ decreases by $n^2$ times.

\par
Solution (\ref{T2}) exhibits three regimes. In the white noise
regime where $\tau_c\Delta \ll 1$, the fluctuating field behaves
like an independent force on the spin precession. The spin
precession angle undergoes a random walk around its mean rate
$d\alpha/dt=E$, and $T_2\approx{\pi^2\over 4\Delta}$, insensitive to
$\tau_c$. In the long correlation regime where the spin experience a
random persistent force, the deviation of the precession angle
increases quadratically with time, and the motion of the spin
behaves like a biased random walk. Correspondingly, decoherence time
$T_2$ is $T_2\approx\pi \sqrt{\tau_c/ (2\Delta)}\ll \tau_c$. In the
intermediate regime of $\tau_c\Delta\sim 1$, the $T_2$, $\tau_c$ and
$\Delta^{-1}$ are in the same order. Unlike the off-diagonal
coupling that affects the spin relaxation most when it is at the
resonance, all fluctuations of the diagonal coupling contribute to
the deviation of the precession angle as it is shown in Eq.
(\ref{deviation}). Thus, no matter whether there are fast (white
noise) or slow (long correlation time) fluctuations, $K_z$ will lead
to an effective decoherence. Of course, it does not mean that all
fluctuations will play an equal role.

\par
Our analysis so far does not rely on the sources of the fluctuating
field. Thus, we expect our results to be applicable to all two level
systems as long as they can be described by a Hamiltonian similar to
(\ref{hamiltonian}). Although $T_1^*$ in our toy-model is more a
definition in a thought experiment than a measure of a true spin
relaxation, it makes the point that the relaxation time for a single
spin and for an ensemble of spins could be totally different. We
present an extreme case where $T_1^*$ is finite while $T_1$, $T_2$
and $T_2^*$ are infinite. In many realistic two-level systems, one
may find another extreme where the distinction between $T_1$ and
$T_1^*$ is not necessary\cite{hu}, but the existence of realistic
systems between the two extremes so that one has to use two
relaxation time to separately describe the relaxation for a single
spin and for an ensemble of spins, just as $T_2$ and $T_2^*$ for
spin decoherence, cannot be ruled out\cite{dassarma,hu,glazman}.

It may be proper to explain the processes that contribute to $T_1$
($T_1^*$) and $T_2$ ($T_2^*$). $T_1$ and $T_1^*$ are caused by the
transitions of a system from one quasi energy level to another. Such
a transition is usually irreversible, accompanying a complete change
of the wave function. This microscopic view of $T_1$, $T_1^*$ is not
in contradiction to the thermodynamic view that $T_1$ is the time of
thermal equilibration for spin population\cite{dassarma} because the
time for a system to reach its thermal equilibrium normally requires
many jumps between the upper and lower levels. Of course, the
numbers in the two definitions may be different, but they should be
in the same order. This is similar to the definition of the electron
relaxation time scattered by impurities, which is the characteristic
time for the electron distribution to return to its equilibrium one;
on the other hand, it can be calculated from the Fermi golden rule
at the quantum mechanical level. $T_2$ and $T_2^*$ come from the
loss of coherence of a state which could be due to a transition for
$T_1$ ($T_1^*$), but it could also due to the accumulation of random
phases to the components of a superposition state without a
transition between two levels, as we showed in our toy-models. Thus,
in general, $T_2$ and $T_2^*$ must be smaller than $T_1$ and
$T_1^*$. In our extreme case, the $T_1^*$ is not caused by true
irreversible transitions, but by coherent transitions (Rabi
oscillations). In this sense, $T_1^*$ in our example is a
pseudorelaxation time, and it should not be surprising to see our
$T_1^*$ deviating from $T_2 < T_1^*$. There are misconceptions in
the literatures about the difference of $T_2$ and $T_2^*$. It was
claimed that $T_2^*$ is referred to as the reversible
dephasing\cite{dassarma}, while irreversible dephasing was
attributed to $T_2$. This claim is questionable because the nature
of randomness is irreversible. Thus both $T_2$ and $T_2^*$ are due
to the irreversible loss of coherence. If the irreversibility in the
claim is referred to the transition from one energy level to
another, it is again not true because both $T_2$ and $T_2^*$ can be
caused by the accumulation of random phases on the expansion
coefficients, as revealed in our second toy-model.

Due to the different natures of $T_1$ and $T_2$, the diagonal and
off-diagonal coupling play quite different roles on the spin
relaxation and the spin decoherence. First, with only diagonal
coupling, there is no spin relaxation. But there is a spin
decoherence. However, a diagonal coupling can also influence the
spin relaxation if an off-diagonal coupling is present. For an
off-diagonal coupling, those fluctuation modes near the Rabi
resonance affects spin relaxation more. Diagonal coupling could
affect the relaxation either through the shifting of resonance point
or through the modification of quantum interference by the phase
change of wave function.

It may also be useful to comment on the relationships between $T_2$
and $T_2^*$. For a non-interacting spin system subject to the same
fluctuating field, we should have $T_2^*< T_2$ because, in addition
to the causes of spin decoherence on individual spin, the spatial
inhomogeneity can contribute to decoherence of an ensemble of spins.
The importance of the inhomogeneity on the decoherence was known in
many fields and also in early work in semiconductor
community,\cite{dassarma,wu} in which the interference effect due to
the inhomogeneity of k space or of real space can cause decay. For
an interacting spin system, it is more complicated. An interaction
can be an extra source of a fluctuating field. On the other hand,the
interaction may glue different spins together. The bondings among
spins could be so strong that all spins have to move coherently. In
this case, one can expect that the dynamics of the system slows
down. Also, a system with a single spin and a system with many spins
are totally different, and there is no comparison. For example, the
spin relaxation and spin decoherence of a Co atom is completely
different from that of a bulk Co magnet where all spins are aligned
in the same direction. And it is not surprising to have spin
decoherence $T_2^*$ of the Co-magnet to be much longer than $T_2$ of
one Co-atom. Another example may be superconductor caused by the
effective e-e attraction. It is known that coherence time of
electrons in a superconducting state could be much longer than that
of an electron in a metal.

In conclusion, we demonstrate the necessity of introducing an
ensemble spin relaxation time $T_1^*$ that could be orders of
magnitude different from $T_1$. We have also shown that all
fluctuations should be relevant to the spin decoherence. There is no
too-slow or too-fast fluctuation for $T_2$ and $T_2^*$.

This work is supported by UGC, Hong Kong, through RGC CERG grants.

\par

\end{document}